\documentclass[11pt,a4paper,twoside]{article}
\pdfoutput=1 
\usepackage{color,graphicx,natbib,lscape,here,geometry,setspace,multirow,enumitem}
\usepackage[dvipsnames]{xcolor}
\usepackage{graphicx}

\usepackage{authblk}
\usepackage{silence}
\usepackage{rotating}
\usepackage{multirow}

\usepackage{booktabs}
\usepackage{color}
\usepackage{setspace}
\usepackage{algorithm2e}
\usepackage{enumitem}
\usepackage{tikz}

\usepackage{newtxtext,newtxmath}
\newcommand*{\scfont}{\fontfamily{ptm}\selectfont}

\definecolor{nblue}{HTML}{000660}
\usepackage[colorlinks=true,urlcolor=nblue,linkcolor=nblue,citecolor=nblue]{hyperref}
\usepackage{bigstrut}

\usepackage{tabularx}
\usepackage{threeparttable}
\usepackage{dcolumn}

\usepackage{etoolbox} 
\makeatletter
\patchcmd{\BR@backref}{\newblock}{\newblock[}{}{}
\patchcmd{\BR@backref}{\par}{]\par}{}{}
\makeatother

\usepackage{pdflscape}
\usepackage{afterpage}

\usepackage{longtable}
\usepackage{array}
\newcolumntype{C}[1]{>{\centering\arraybackslash}p{#1}}

\usepackage{comment}
\usepackage{mathtools}

\usepackage[hang,flushmargin]{footmisc} 
\usepackage[british]{babel}

\pdfoutput=1
 \usepackage{float}
 \restylefloat{table}

\usepackage[title,titletoc]{appendix}
\makeatletter

\renewenvironment{appendices}{%
    \begin{oldappendices}%
    \renewcommand{\thefigure}{\ifnum \c@section>\z@ \thesection.\fi\@arabic\c@figure}%
    \@addtoreset{figure}{section}%
    \renewcommand{\thetable}{\ifnum \c@section>\z@ \thesection.\fi\@arabic\c@table}%
    \@addtoreset{table}{section}}{%
    \end{oldappendices}%
}\makeatother

\usepackage{titlesec} 
\titleformat{\section}[block]{\large}{\thesection. }{0em}{\MakeUppercase} 
\titleformat{\subsection}[block]{\large}{\thesubsection. }{0em}{\itshape} 
\titleformat{\subsubsection}[block]{\large}{}{0em}{\itshape} 

\let\natbibcitet\citet
\renewcommand\citet{\bibpunct{(}{)}{,}{a}{,}{,}\natbibcitet}
\let\natbibcitep\citep
\renewcommand\citep{\bibpunct{(}{)}{;}{a}{,}{;}\natbibcitep}
\newcommand{\bi}{\begin{itemize}}
\newcommand{\ei}{\end{itemize}}
\newcommand{\be}{\begin{equation}}
\newcommand{\ee}{\end{equation}}
\defcitealias{ieo14}{IEO, 2014}

\long\def\symbolfootnote[#1]#2{\begingroup%
\def\thefootnote{\fnsymbol{footnote}}\footnote[#1]{#2}\endgroup}

\makeatletter
\def\ubar#1{\underline{\sbox\tw@{$#1$}\dp\tw@\z@\box\tw@}}
\def\obar#1{\overline{\sbox\tw@{$#1$}\dp\tw@\z@\box\tw@}}
\makeatother

\usepackage{bm}
\usepackage{caption}
\usepackage{subcaption}

\makeatletter\let\p@subfigure\thefigure\makeatother

\floatstyle{plaintop}
\restylefloat{table}

\usepackage{fancyhdr} 
\usepackage{cleveref}
\crefname{chapter}{Chapter}{Chapters}
\crefname{section}{Section}{Sections}
\crefname{subsection}{Section}{Sections}
\crefname{subsubsection}{Section}{Sections}
\crefname{figure}{Figure}{Figures}
\crefname{table}{Table}{Tables}
\crefname{equation}{Equation}{Equations}
\crefname{appendix}{Appendix}{Appendices}
\crefname{appendices}{Appendix}{Appendices}

\crefname{appsec}{Appendix}{Appendices}

\usepackage{catoptions}
\makeatletter

\def\Autoref#1{%
  \begingroup
  \edef\reserved@a{\cpttrimspaces{#1}}%
  \ifcsndefTF{r@#1}{%
    \xaftercsname{\expandafter\testreftype\@fourthoffive}
      {r@\reserved@a}.\\{#1}%
  }{%
    \ref{#1}%
  }%
  \endgroup
}
\def\testreftype#1.#2\\#3{%
  \ifcsndefTF{#1autorefname}{%
    \def\reserved@a##1##2\@nil{%
      \uppercase{\def\ref@name{##1}}%
      \csn@edef{#1autorefname}{\ref@name##2}%
      \autoref{#3}%
    }%
    \reserved@a#1\@nil
  }{%
    \autoref{#3}%
  }%
}
\makeatother

\newcolumntype{d}[1]{D{.}{.}{#1}}

\title{\huge{Stochastic model specification in Markov switching vector error correction models}}
\author{\large{\uppercase{Niko Hauzenberger}$^{1,2}$, \uppercase{Florian Huber}$^{1}$, \uppercase{Michael Pfarrhofer}$^{1,2}$ \\\vspace*{-0.75em} and \uppercase{Thomas O. Z\"{o}rner}$^{2}$\thanks{Corresponding author: Thomas O. Z\"{o}rner. E-mail: \href{mailto:tzoerner@wu.ac.at}{tzoerner@wu.ac.at}. The authors acknowledge funding from the Austrian Science Fund (FWF) for the project ``High-dimensional statistical learning: New methods to advance economic and sustainability policies'' (ZK 35), jointly carried out by WU Vienna University of Economics and Business, Paris Lodron University Salzburg, TU Wien, and the Austrian Institute of Economic Research (WIFO), and financial support from the Austrian National Bank, Jubilaeumsfond grant no. 17650.}}\\\vspace*{-0.5em}\normalsize{
$^1$\textit{Paris Lodron University of Salzburg}\\
$^2$\textit{Vienna University of Economics and Business}}}
\date{}


\def\equationautorefname~#1\null{%
  Eq.~(#1)\null
}
\def\equationautorefname~#1\null{
Eq.~(#1)\null
}

\usepackage[utf8, latin1]{inputenc}                 
\begin{document}
\maketitle\thispagestyle{empty}\normalsize\vspace*{-2em}\small

\begin{center}
\begin{minipage}{0.8\textwidth}
\noindent\small This paper proposes a hierarchical modeling approach to perform stochastic model specification in Markov switching vector error correction models. We assume that a common distribution gives rise to the regime-specific regression coefficients. The mean as well as the variances of this distribution are treated as fully stochastic and suitable shrinkage priors are used. These shrinkage priors enable to assess which coefficients differ across regimes in a flexible manner. In the case of similar coefficients, our model pushes the respective regions of the parameter space towards the common distribution. This allows for selecting a parsimonious model while still maintaining sufficient flexibility to control for sudden shifts in the parameters, if necessary. We apply our modeling approach to real-time Euro area data and assume transition probabilities between expansionary and recessionary regimes to be driven by the cointegration errors. The results suggest that the regime allocation is governed by a subset of short-run adjustment coefficients and regime-specific variance-covariance matrices. These findings are complemented by an out-of-sample forecast exercise, illustrating the advantages of the model for predicting Euro area inflation in real time. 
\end{minipage}
\end{center}
\bigskip\normalsize\clearpage

\renewcommand{\thepage}{\arabic{page}}

\section{Introduction}
In this paper we propose a hierarchical multivariate regime switching model that allows for detecting which parameters differ across regimes. The model, a Markov switching vector error correction model (MS-VECM), explicitly discriminates between short- and long-run dynamics and potentially allows for time-varying transition probabilities that depend on the cointegration errors. We assess whether parameters differ across regimes by using novel shrinkage priors that have recently been utilized in finite mixture modeling \citep[see][]{yau2011hierarchical, malsiner2016model}. 

The literature on Bayesian estimation of Markov switching (MS) models is voluminous \citep[see, among many others,][]{chib1996calculating, kim1998business, kaufmann2000measuring, sims2008methods, kaufmann2015k, droumaguet2017granger, boganni2018msvar}. By contrast, contributions that explicitly deal with the Bayesian estimation of MS-VECMs are comparatively sparse \citep[][]{martin2000us,paap2003bayes,jochmann2015regime}. Most of these contributions use Bayesian shrinkage priors to enable reliable and efficient estimation. These priors are typically specified in the spirit of standard Minnesota priors \citep[][]{doan1984forecasting, sims1998bayesian} and symmetric across regimes. Here, one implication is that coefficients are pushed towards a stylized prior model (like a multivariate random walk), irrespective of the regime and what the (regime-specific) likelihood suggests. For instance, a regression parameter may be zero in one regime, but different in another. Such a situation is effectively ruled out as both coefficients are pushed towards zero.

In this paper, we circumvent such issues by proposing a novel hierarchical modeling approach that has originally been proposed in the literature on finite mixture models \citep[see][]{yau2011hierarchical, malsiner2016model}. We estimate an MS-VECM assuming that the regime-specific coefficients arise from a common distribution. The mean of this common distribution is treated as unknown and estimated from the data.  Using Normal-Gamma  shrinkage priors \citep[][]{griffin2010inference} on the variance-covariance matrix of the common distribution  enables us to gain an understanding on what covariates drive the corresponding regime allocation. When compared to the existing literature, our approach allows for flexibly testing which coefficients (or sets of them) should differ across regimes while pushing similar coefficients towards a common mean. In addition, we follow the literature on MS models with time-varying transition probabilities \citep[see, among others,][]{filardo1994business, kim1998business, kaufmann2015k} and assume that the transition probability matrix is time-varying and depends on the lagged cointegration errors. This setup implies that we test whether deviations of macroeconomic fundamentals from their long-run equilibrium values impact the transition probabilities between business cycle stages.

For an empirical illustration, we estimate a medium-scale Euro area business cycle model that discriminates between business cycle expansions and contractions based on real-time data \citep{giannone2012area}. The results suggest that our model successfully replicates Euro area business cycle behavior and performs well in forecasting inflation relative to a set of competing specifications. We find that deviations of inflation, unemployment and expected inflation (measured by the 2-year inflation projections of the ECBs survey of professional forecasters) from their long-run fundamentals have predictive power for the transition probabilities. When considering differences in parameters across regimes, we find that some short-run adjustment coefficients differ markedly across regimes. Moreover, a subset of autoregressive coefficients displays variation across regimes, while others exhibit no structural breaks.  Besides allowing for time variation in the autoregressive coefficients, it appears crucial to consider heteroscedastic shocks by allowing for a regime-specific variance-covariance matrix.

The remainder of the paper is organized as follows. Section 2 outlines the proposed idea by means of a simple  switching regression model while Section 3 presents the  MS-VECM model as well as the prior setup. Section 4 first gives an overview of the dataset used and subsequently shows the empirical results and the forecasting performance over time. The last section summarizes and concludes the paper.

\section{A simple hierarchical model}\label{sec: simpleEX}
In this section we outline the main idea on how to determine whether coefficients differ across regimes within a simple regression model. Subsequently, we generalize the stylized example to a multivariate nonlinear error correction model. 

We set the stage by assuming that a scalar time series $\{y\}_{t=1}^T$ follows a switching regression model \citep[][]{goldfeld1973markov},\footnote{For textbook introductions, see \cite{kim1999state} and \cite{fruhwirth2006finite}.}
\begin{equation}
y_t = \beta_{1S_t} x_{1t} + \beta_{2 S_t} x_{2t}+ \sigma_{S_t} \eta_t, \quad \eta_t \sim \mathcal{N}(0,1),
\end{equation}
with $x_{jt}$ being exogenous covariates and $\beta_{j S_t}$ (for $j=1,2$) the associated regression coefficients while $\sigma^2_{S_t}$ denotes the error variance. We assume that $S_t$ represents a latent quantity that takes values $0,\dots, R$ and follows a first-order Markov process.\footnote{The arguments we provide below hold for any law of motion of $S_t$.}

The Bayesian literature typically uses Gaussian priors on $\beta_{j S_t}$,
\begin{equation}
\beta_{j S_t} \sim \mathcal{N}(0, \tau_j), \label{eq: prior_0}
\end{equation}
where $\tau_j$ denotes a fixed prior scaling for coefficient $j$. Notice that $\tau_j$ is regime invariant and the prior is centered on zero. In time series applications and for non-stationary data, it is common practice that the prior on the first lag of $y_t$ is centered on unity, while for higher lag orders the prior mean is set equal to zero. 

The symmetric prior in \autoref{eq: prior_0} translates into a prior on the standardized distance between coefficients.  For illustration, we compute the prior distance between $\beta_{j k}$ and $\beta_{j l}$ for $k \neq l$,
\begin{equation}
\frac{\beta_{j k} - \beta_{j l}}{\sqrt{2}} \sim \mathcal{N}(0, \tau_j). \label{eq: distance}
\end{equation}
If $\tau_j$ is close to zero, the corresponding coefficients do not differ significantly across regimes. However, they are simultaneously strongly shrunk to zero. Thus, while such a prior is able to push selected coefficients towards homogeneity, it is not capable of handling cases where coefficients are non-zero but at the same time appear to be the same across regimes.

As a solution, we follow \cite{yau2011hierarchical} and assume that the regime-specific coefficients arise from a common distribution,
\begin{equation}
\beta_{j S_t} \sim \mathcal{N}(\beta_j, \tau_j). \label{eq: prior_1}
\end{equation}
The common mean $\beta_j$ is treated as unknown and estimated from the data.  It is noteworthy that if $\tau_j$ is close to zero, $\beta_{j S_t}$ is pushed to $\beta_j$ across all regimes. Computing the standardized distance between $\beta_{j k}$ and $\beta_{j l}$ yields \autoref{eq: distance} and the same intuition applies. However, instead of pulling $\beta_{j S_t}$ to zero for all regimes, this specification shrinks towards parameter homogeneity and \autoref{eq: prior_1} can be interpreted as a Gaussian hierarchical prior on $\beta_{j S_t}$.

In what follows, we use shrinkage priors on $\tau_j$ to test the existence of significant differences across regimes. Specifically, we assume that $\tau_j$ arises from a Gamma distribution,
\begin{equation}
\tau_j \sim \mathcal{G}(d_0, d_1), \label{eq: priorgamma}
\end{equation}
with $d_0$ and $d_1$ being hyperparameters. This specification has been introduced by \cite{griffin2010inference} within a regression context and subsequently adopted by \cite{malsiner2016model} to determine variable relevance in finite mixture models. Using a Gamma prior allows for shrinking $\tau_j$ to zero if necessary, and thus permits selecting whether parameters differ across regimes. 

\section{Econometric framework}
In Section \ref{sec:msvecm}, we first discuss the multivariate MS error correction specification while Section \ref{sec:priors} discusses the prior choice and Section \ref{sec:posteriors} briefly summarizes the main steps necessary to perform posterior inference.

\subsection{A nonlinear vector error correction model}\label{sec:msvecm}
The simple model outlined in the previous section is now generalized to a MS-VECM with two regimes. We opt for two regimes since we are interested in developing a model of the Euro area that encompasses several stylized facts about business cycles, namely pronounced  co-movement of macroeconomic quantities over the business cycle as well as the distinction between expansionary and recessionary stages \citep[][]{burns1946measuring}. Moreover, our proposed framework explicitly aims at discriminating between short- and long-run dynamics.

We assume that the first difference of an $m$-dimensional vector of time series $\{\bm y_t\}_{t=1}^T$ follows an MS-VECM with $P$ lags,
\begin{equation}
 \Delta \bm y_t = \bm \lambda_{S_t} \bm b' \bm y_{t-1}+\sum_{p=1}^P \bm B_{p S_t}  \Delta \bm y_{t-p} + \bm H_{S_t} \bm \eta_t, \quad \bm \eta_t \sim \mathcal{N}(\bm 0 , \bm I_m). \label{eq: tvecm}
\end{equation}
Here we let $\bm \lambda_{S_t}$ be an $m \times r$ matrix of short-run adjustment coefficients, $\bm b$ is an $m \times r$ matrix of long-run relations and $\bm B_{p S_t}$ are $m \times m$ coefficient matrices associated with the $p$th lag of $\Delta \bm y_t$. Furthermore, $\bm H_{S_t}$ is the lower Cholesky factor of a regime-specific variance-covariance matrix $\bm \Sigma_{S_t}= \bm H_{S_t} \bm H'_{S_t}$ and $S_t$ is a discrete Markov process that takes values zero or unity detailed below.\footnote{Note that it is also possible to estimate different chains for the coefficients and volatilities, as in \citet{HUBRICH2015100}.} In what follows we rewrite \autoref{eq: tvecm} as 
\begin{equation}
\Delta \bm y_t = \bm A_{S_t} \bm x_t +  \bm H_{S_t} \bm \eta_t, \label{eq: tvecm_reg}
\end{equation}
whereby $\bm A_{S_t} = (\bm \lambda_{S_t}, \bm B_{1 S_t}, \dots, \bm B_{P S_t})$ is an $m \times K$ matrix of stacked  coefficients with $K= r+ m P$. Moreover,   $\bm x_t = (\bm w'_t, \Delta \bm y'_{t-1}, \dots, \Delta \bm y'_{t-P})'$ and $\bm w_t = \bm b' \bm y_{t-1}$ denotes the $r$ cointegration errors.\footnote{For notational simplicity we suppress the dependence of $\bm w_t$ on $\bm b$.}  Notice that \autoref{eq: tvecm_reg} is a standard multivariate regression model conditional on $\bm b$. We explicitly rule out the possibility of breaks in $\bm b$ since they are assumed to be long-run fundamental relations and thus not subject to abrupt changes. However, there exist studies which allow for nonlinearities in the cointegration relationship \citep[see, for instance, ][]{martin2000us, rahbek2004, koop2011bayesian, jochmann2015regime}. 

The transition probabilities of $S_t$,  $\Pr(S_t = j | S_{t-1} = i, \bm \gamma, \bm w_t)=p_{ij, t}$, are specified to be time-varying and depend on the (lagged) cointegration errors $\bm w_t$ through a set of regression coefficients in $\bm{\gamma}$. More precisely, the matrix of transition probabilities is given by
\begin{equation}
\boldsymbol{P}_t = \begin{pmatrix}
p_{00,t} & p_{01,t} \\
p_{10,t} & p_{11,t}
\end{pmatrix}. \label{eq: transProb}
\end{equation}
The rows of \autoref{eq: transProb} sum to unity for all $i$ and $t$.  Following the literature on early warning Markov switching models \citep[][]{filardo1994business, kim1998business, amisano2013money, huber2018markov}, we assume that the transition probabilities are parameterized using a probit specification,\footnote{In the case of more than two regimes, a potential alternative would be a logit specification \citep[][]{kaufmann2015k, billio2016interconnections}.}
\begin{align}
p_{ij,t}= \Phi ( c_{0 i}+\bm{\gamma}' \boldsymbol{w}_{t}),
\label{eq:transprob}
\end{align}
with $c_{0i}$ denoting a regime-specific intercept term and $\Phi$ denotes the cumulative distribution function of the standard normal distribution. The $j$th element in $\bm \gamma$ measures the sensitivity of the transition probabilities with respect to the $j$th cointegration error, $w_{jt}$. Following \cite{amisano2013money}, we postulate that $\bm \gamma$ is time-invariant  while the intercept term depends on the prevailing regime. Using the latent variable representation of the probit model yields
\begin{equation}
z_t^* = c_{0 i} + \bm {\gamma}' \bm w_t + u_t, \quad u_t \sim \mathcal{N}(0, 1).
\end{equation}
For identification purposes, the error variance is set equal to unity.

Before proceeding to the prior specification it is worth noting that  $\bm \lambda$ and $\bm b$  are not identified since they enter \autoref{eq: tvecm} as a product. We achieve identification by using the linear normalization $\bm b = (\bm I_r, \bm \Xi')' $ with $\bm \Xi$ being a $(m-r) \times r$ matrix of coefficients. This choice is clearly not invariant to the ordering of the elements $\bm y_t$ but ensures that the model is exactly identified.\footnote{Another potential choice  would be to identify the space spanned by the cointegrating vectors and introduce a restriction on this space \citep[][]{strachan2003valid, koop2009efficient}.}

\subsection{Prior specification}\label{sec:priors}
The model outlined in the previous section is heavily parameterized and we thus adopt a Bayesian approach to estimation and inference. Consistent with the discussion in Section \ref{sec: simpleEX} we assume that  the common distribution that gives rise to $\bm a_{S_t}= \text{vec}(\bm A_{S_t})$ follows a multivariate Gaussian distribution,
\begin{equation}
\bm a_{S_t} \sim \mathcal{N}(\bm a, \bm \Omega), \label{eq: common_dist}
\end{equation}
where $\bm \Omega = \text{diag}(\tau_1, \dots, \tau_k)$ is a diagonal variance-covariance matrix with variances $\tau_j$ for $k = Km$. Analogous to Section \ref{sec: simpleEX}, $\tau_j$ determines the similarity between elements in $\bm a_{0}$ and $\bm a_1$. For instance, if only the short-run adjustment coefficients differ across regimes, the corresponding elements in $\bm \Omega$ will be rather large whereas for the remaining coefficients, the associated $\tau_j$s will be close to zero.

Following \cite{malsiner2016model} we specify a Gaussian prior on $\bm a$ and the Gamma prior outlined in \autoref{eq: priorgamma} for all $j$,
\begin{align}
\bm a &\sim \mathcal{N}(\underline{\bm a}, \underline{\bm \Omega}),\\
\tau_j &\sim \mathcal{G}(d_0, d_1) \label{eq: shrink_TAU}. 
\end{align}
Hereby, we let $\underline{\bm a}$ denote the $k$-dimensional prior mean vector and $\underline{\bm \Omega}$ is a $k\times k$  prior variance-covariance matrix. In what follows, we set $\underline{\bm a}= \bm 0$  and $\underline{\bm \Omega}^{-1} = 0 \times \bm I_k$, leading to an improper prior  on $\bm a$.  For $\tau_j$, it is worth emphasizing that if $d_0=1$, we obtain the Bayesian Lasso \citep[][]{park2008bayesian} used in \cite{yau2011hierarchical} while if $d_0 <1$ increasing prior mass is placed on zero while the tails of the marginal prior become heavier \citep[][]{malsiner2016model}. In our empirical application, we set $d_0 = d_1=0.1$ to strongly center the prior on zero and allow for heavy tails.

The prior setup described in Eqs. (\ref{eq: common_dist})--(\ref{eq: shrink_TAU}) effectively allows for detecting which elements in $\bm a_{S_t}$  differ across regimes and which of them appear to be homogeneous over distinct business cycle stages. Especially in light of a moderate to large number of time series in $\bm y_t$ as well as a moderate number of lags $P$, the number of parameters per regime is large relative to the length of a typical dataset. In the presence of multiple regimes, however, this problem is even more severe and shrinkage is necessary to obtain reliable parameter estimates.

It is worth mentioning that the proposed setup suffers from the well-known label switching problem \citep[see][]{fruhwirth2006finite}. This issue arises due to the labeling of the states not affecting the likelihood function. In such cases, one may either rely on random permutation techniques, or achieve identification by employing economic reasoning. In the empirical application, for instance, we assume that the conditional mean of the equation for unemployment is lower for $S_t=0$, and achieve this by implementing a rejection step in the MCMC algorithm.

We proceed with the prior on the free elements of the long-run adjustment coefficients. For $\bm \xi= \text{vec}(\bm \Xi)$, the $v=(m-r) r$ free elements of $\bm b$, we use a Gaussian prior,
\begin{equation}
\bm \xi \sim \mathcal{N}(\bm 0 , \zeta \times \bm I_v),
\end{equation}
where $\zeta$ is a prior hyperparameter that controls the tightness of the prior. We set $\zeta=1$, which is a fairly uninformative choice given the scale of our data. In principle, it would also be possible to elicit a prior directly on the cointegrating space \citep[][]{strachan2003valid}. Here, we follow the traditional approach since we are interested in directly interpreting the corresponding cointegration error.

Following \cite{fruhwirth2006finite}, \cite{malsiner2016model} and \cite{huberzoernerTVECM}, we use a hierarchical  Wishart  prior on $\bm \Sigma^{-1}_{S_t}$,
\begin{align}
\bm \Sigma^{-1}_{S_t} &\sim \mathcal{W}( \bm S,  s), \\
\bm S &\sim \mathcal{W}(\bm Q,  q).
\end{align}
The prior hyperparameters are specified such that \citep[see][]{fruhwirth2006finite, malsiner2016model}
\begin{align}
s &= 2.5 + \frac{m-1}{2},\\
q &= 0.5 +\frac{m-1}{2},\\
\bm Q &=  \frac{100 s}{\underline{s}} ~\text{diag}(\hat{\sigma}^2_1, \dots, \hat{\sigma}^2_m).
\end{align}
Here we let $\hat{\sigma}^2_j$ denote the OLS variance obtained by running $m$ univariate autoregressive models of order $P$. This choice implies that the all of the regime-specific variance-covariance matrices stem from a common distribution, similar to the case for $\bm a_{S_t}$ outlined above.

Finally, we use Gaussian priors on $\bm \Gamma = (c_{0 i} , \bm \gamma')'$,
\begin{equation}
\bm \Gamma \sim \mathcal{N}(\bm 0 , \bm V),
\end{equation}
with $\bm V = 10 \times \bm I_{r+1}$ to introduce relatively little prior information on $\bm \Gamma$. Decreasing the prior variance would lead to a situation where the researcher suspects that transition probabilities do not depend on $\bm w_t$ and tend to be time-invariant.

\subsection{Full conditional posterior simulation}\label{sec:posteriors}
The model outlined in the previous sections features a joint posterior density that is intractable. Fortunately, all full conditional posterior distributions take a well known form and are thus amenable to perform Gibbs sampling. In this section, we briefly summarize the steps involved in order to obtain a valid draw from the joint posterior, focusing attention on the non-standard parts while providing references for the full conditionals that are standard.

Conditional on a suitable set of starting values, our Markov chain Monte Carlo (MCMC) algorithm cycles through the following steps:
\begin{enumerate}
\item Sample $\bm a_{j}~(j=0,1)$ conditional on the remaining parameters and the states from a $k$-dimensional multivariate Gaussian distribution. The corresponding moments take a standard form \citep[][]{zellner1996introduction}.
\item Simulate $\bm \xi$  from a Gaussian posterior distribution conditional on the remaining parameters, states and a set of identifying assumptions. The precise formulas can be found in \cite{villani2001bayesian} and \cite{huberzoernerTVECM}.
\item The common mean $\bm a$ is simulated conditional on $\bm a_0, \bm a_1$ and $\underline{\bm \Omega}$ from a Normal distribution, with $\odot$ indicating point-wise multiplication,
\begin{align*}
\bm a | \bm a_0, \bm a_1, \bm \Omega &\sim \mathcal{N}(\overline{\bm a},  \overline{\bm \Omega})\\
\overline{\bm \Omega} &=  \frac{1}{2} \bm \Omega, \\
\overline{\bm a} &=  \frac{1}{2}~ \left(\bm a_0 + \bm a_1\right).
\end{align*}
\item Draws from the conditional posterior of $\tau_j$ are obtained by noting that $p(\tau_j | \bm a_0, \bm a_1)$ follows a generalized inverted Gaussian (GIG) distribution,
\begin{equation}
\tau_j | \bm a, \bm a_0, \bm a_1 \sim \mathcal{GIG}\left(d_0 - 1, \sum_{j=0}^1 (\bm a_j - \bm a)^2, 2 d_1\right).
\end{equation}
\item Draw $\bm \Sigma^{-1}_{j}$ (for $j=0,1$) from a Wishart conditional posterior distribution given by
\begin{align*}
\bm \Sigma^{-1}_{j}|\bm a_0, \bm a_1, S^T, \bm S, \mathcal{D} &\sim \mathcal{W}(\overline{\bm S}_{j}, \overline{s}_{j})\\
\overline{\bm S}_{j} &= \bm S + \frac{1}{2}\sum_{t: S_t =j} (\Delta \bm y_t - A_{j} \bm x_t')(\Delta \bm y_t - A_{j} \bm x_t')',\\
\overline{s}_{j} &= s+N_j/2
\end{align*}
whereby $S^T = (S_1, \dots, S_T)'$ denotes the full history of the states, $\mathcal{D}$ the data and $N_j = \#(t: S_t =j)$, that is, the number of observations in regime $j$.
\item The common scaling matrix for the Wishart prior $\bm S$ is simulated from its Wishart distributed conditional distribution,
\begin{align*}
\bm S |\bm a_0, \bm a_1 &\sim \mathcal{W}(\overline{\bm S}, \overline{q})\\
\overline{\bm S} &= \bm Q + \sum_{j=0}^1\bm \Sigma_j^{-1}, \\
\overline{q} &=q+2 s.
\end{align*}
\item Simulate the full history of the states $S^T$ as well as the transition probabilities using the algorithm outlined in \cite{kim1998business} and adopted in \cite{amisano2013money}.

\item Estimate the full history of $z^*_t$ and $\bm \Gamma$ using the algorithm proposed in \cite{albert1993bayesian}.
\end{enumerate}
We repeat this algorithm 85,000 times where the first 50,000 draws are discarded as burn-in. Convergence is assessed using standard trace plots as well as inefficiency factors and the \cite{raftery1992many} diagnostic. All measures point toward rapid convergence for most parts of the parameter space under scrutiny.\footnote{Detailed convergence diagnostics can be obtained from the authors upon request.}

\section{Empirical application}
In this section, we first present our dataset and discuss details on model specification. We proceed by illustrating the features of the modeling approach by providing key in-sample empirical findings. The section is complemented by a real-time out-of-sample forecasting exercise.

\subsection{Data overview and model specification} \label{sec: specification}
For the empirical application we use the real-time database of \citet{giannone2012area}, covering quarterly Euro area time series ranging from $1999$:Q$1$ to $2017$:Q$4$. Based on our focus of modeling inflation dynamics in the forecasting exercise, we select five quantities of interest from the full dataset and include two-year inflation projections of the ECB's survey of professional forecasters to measure inflation expectations \citep[see, for instance,][]{koop2012estimating,jarocinski2018inflation}. 

The information set comprises the growth rate of the harmonized index of consumer prices (excluding energy and unprocessed food, labeled HICPEX), the unemployment rate (UNEMP), inflation expectations (HICP-EXPEC), Euro area capacity utilization in manufacturing (UTILIZ) in logs, oil prices (OIL) in logs, and the three-month Euribor (I3M) as a short-term interest rate. Due to the normalization assumption on $\bm b$, the ordering of the variables in $\bm y_t$ plays a crucial role. Here, we order
\begin{equation}
\bm y_t = (\textit{HICPXE}, \textit{UNEMP}, \textit{HICP-EXPEC},  \textit{UTILIZ}, \textit{OIL}, \textit{I3M})'. \label{eq: ordering}
\end{equation}
With this ordering we assume that the equilibrium realized inflation depends on unemployment, the inflation expectations, the capacity utilization, oil prices and the interest rate, resembling a variant of the expectation augmented Phillips curve. As is standard in the literature for quarterly data, we choose $P = 4$ lags in the empirical specification and include an intercept term. This translates to five lags of $\bm y_t$ in terms of a vector autoregression (VAR) in levels.

\subsection{Posterior distribution of the cointegrating errors}\label{sec: cointrank}

A crucial modeling decision in the context of VECMs is to select the appropriate cointegration rank $r$. Our specific choice of $r$ is based on using log predictive likelihoods, reported in  Section \ref{sec:forecasting}. The following discussion is based on setting $r=3$, which is a choice that makes sense from a theoretical perspective but also yields strong density predictions (see the discussion in Section \ref{sec:forecasting}).

\begin{figure}[!t]
\centering 
\begin{minipage}[t]{\textwidth}
\centering 
\textbf{(a)} First cointegration error
\end{minipage}
\begin{minipage}[t]{\textwidth}
\includegraphics[width=\textwidth]{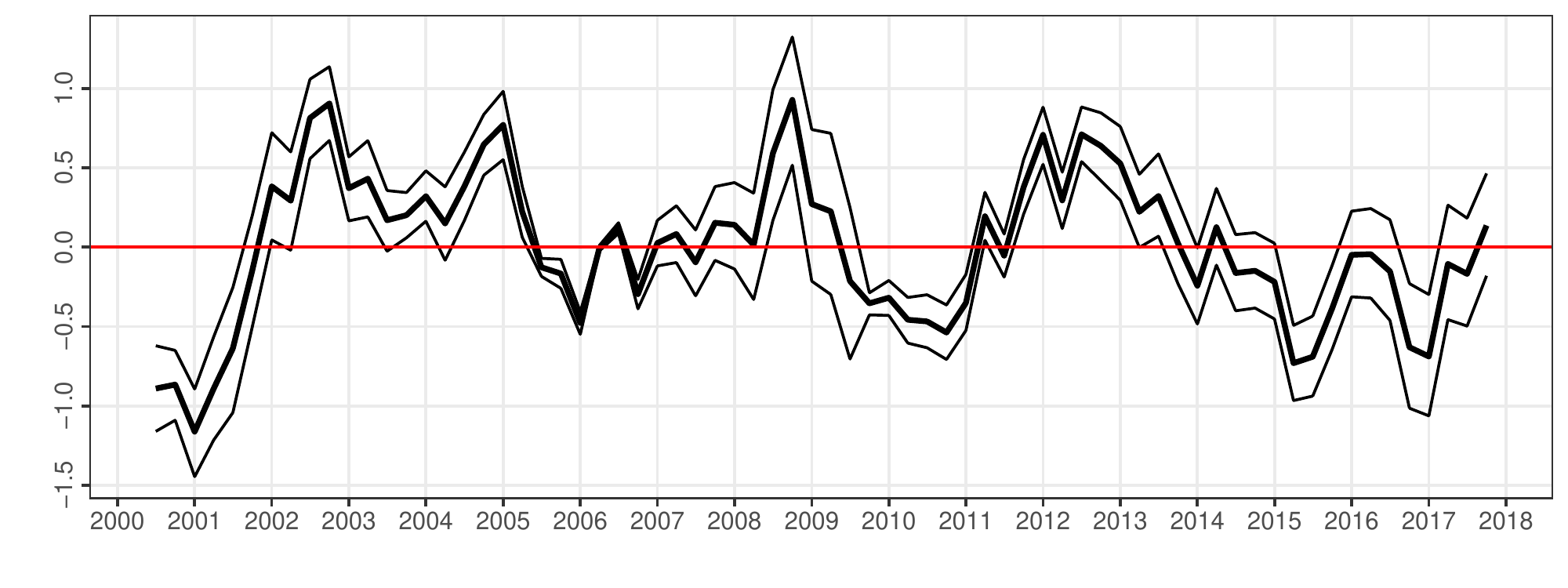}
\end{minipage}
\begin{minipage}[t]{\textwidth}
\centering 
\textbf{(b)} Second cointegration error
\end{minipage}
\begin{minipage}[t]{\textwidth}
\includegraphics[width=\textwidth]{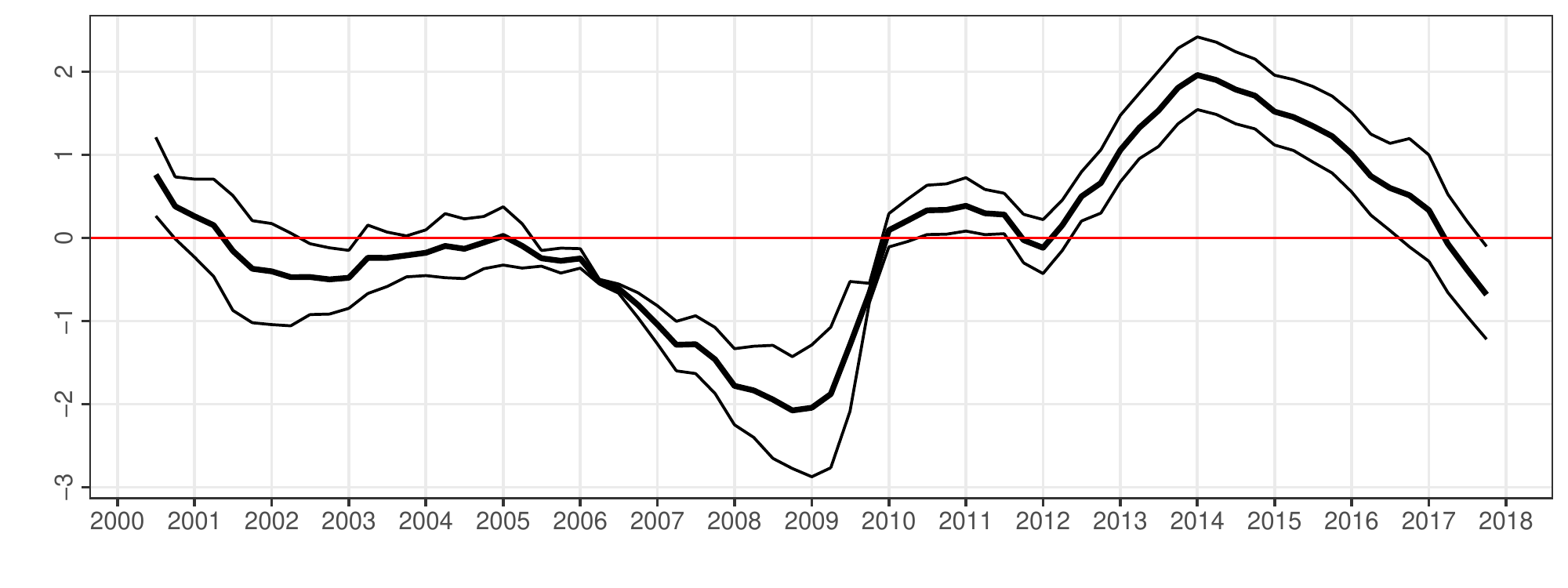}
\end{minipage}
\begin{minipage}[t]{\textwidth}
\centering 
\textbf{(c)} Third cointegration error
\end{minipage}
\begin{minipage}[t]{\textwidth}
\includegraphics[width=1\textwidth]{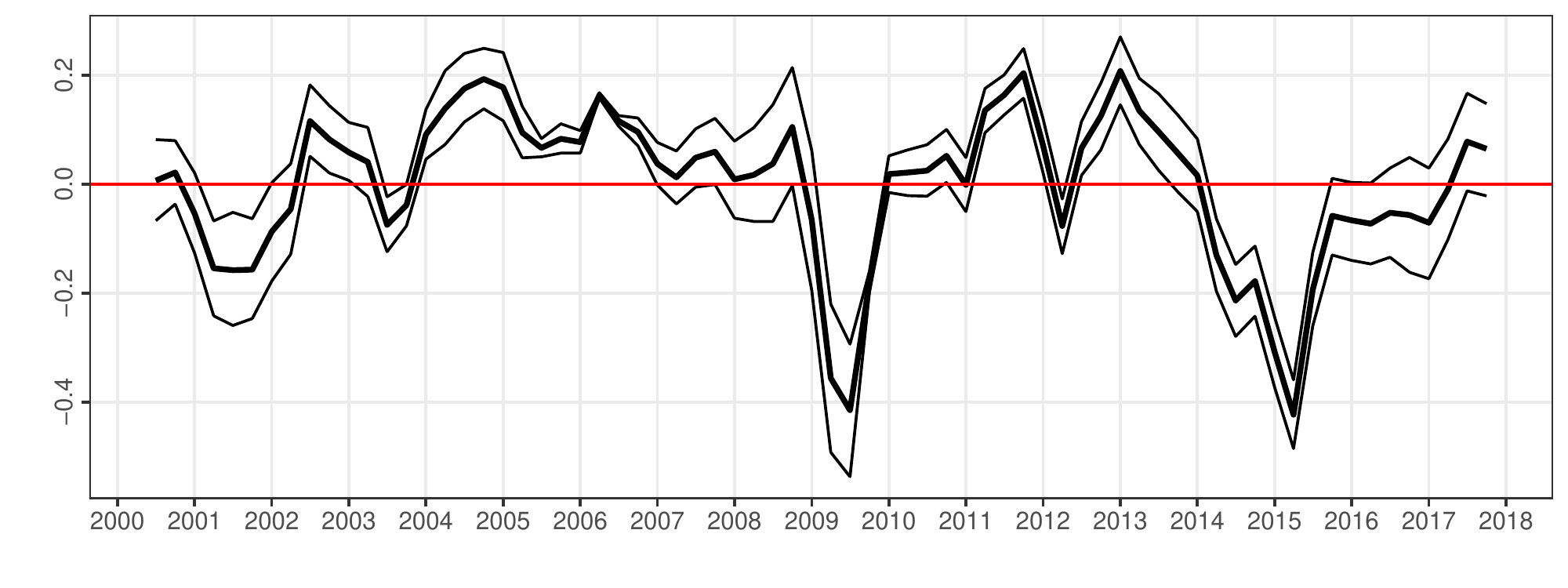}
\end{minipage}
\caption{Posterior distribution of the cointegration errors.}\label{fig: coint_errors}\vspace{-0.3cm}
\caption*{\footnotesize\textit{Notes}: The thick black line is the posterior median of the cointegration errors, while the thin black lines refer to the 16th and 84th percentiles. The red line marks zero.}
\end{figure}

The posterior distributions of the three cointegration errors that are subsequently used to inform the transition distributions are depicted in \autoref{fig: coint_errors}. Notice that the first error can be interpreted as the deviation of inflation from its long-run equilibrium value determined by the remaining elements in $\bm y_t$. Consequently, the second error can be considered as the deviation of unemployment from its long-run equilibrium value, related to the non-accelerating inflation rate of unemployment (NAIRU). By contrast, the third error can be viewed as the deviation of inflation expectations from the target level.

The first cointegration error in \autoref{fig: coint_errors}(a) indicates that inflation is below its equilibrium value during the first two years of the sample. From 2002 onward, it quickly reverts towards its long-run fundamental value and overshoots until mid 2005. The figure suggests that the difference between inflation and its long-run equilibrium value increased until the onset of the European sovereign debt crisis. From this point on, inflation is below its implied equilibrium value in most quarters, except for the period between 2012 and 2013. Turning to the second cointegration error, \autoref{fig: coint_errors}(b) indicates that unemployment is roughly at its long-run equilibrium until the beginning of 2006, while negative errors occur subsequently until 2009. In the build-up to the Great Recession, we observe large negative cointegration errors for unemployment, mirroring conventional measures of the Euro area output gap. A sharp increase occurs in early 2010, and the Double-Dip recession of the Euro area is clearly visible. By the end of 2015, the model suggests a return of unemployment to its long-run equilibrium value. The third cointegration error in \autoref{fig: coint_errors}(c) refers to the deviation of inflation expectations from trend inflation. Three distinct periods are worth emphasizing: First, a period of approximate equilibrium that harshly ended during the financial crisis with strong negative deviations. Second, two pronounced positive spikes are observable between 2011 and 2014. Finally, we observe negative cointegration errors in 2014 and 2015 similar in magnitude to the dip during the Great Recession, while inflation expectations return to their equilibrium level towards the end of the sample period.

\subsection{Regime allocation and time-varying transition distributions}\label{sec: states}
\begin{figure}[!t]
	\centering
	\includegraphics[width=\textwidth]{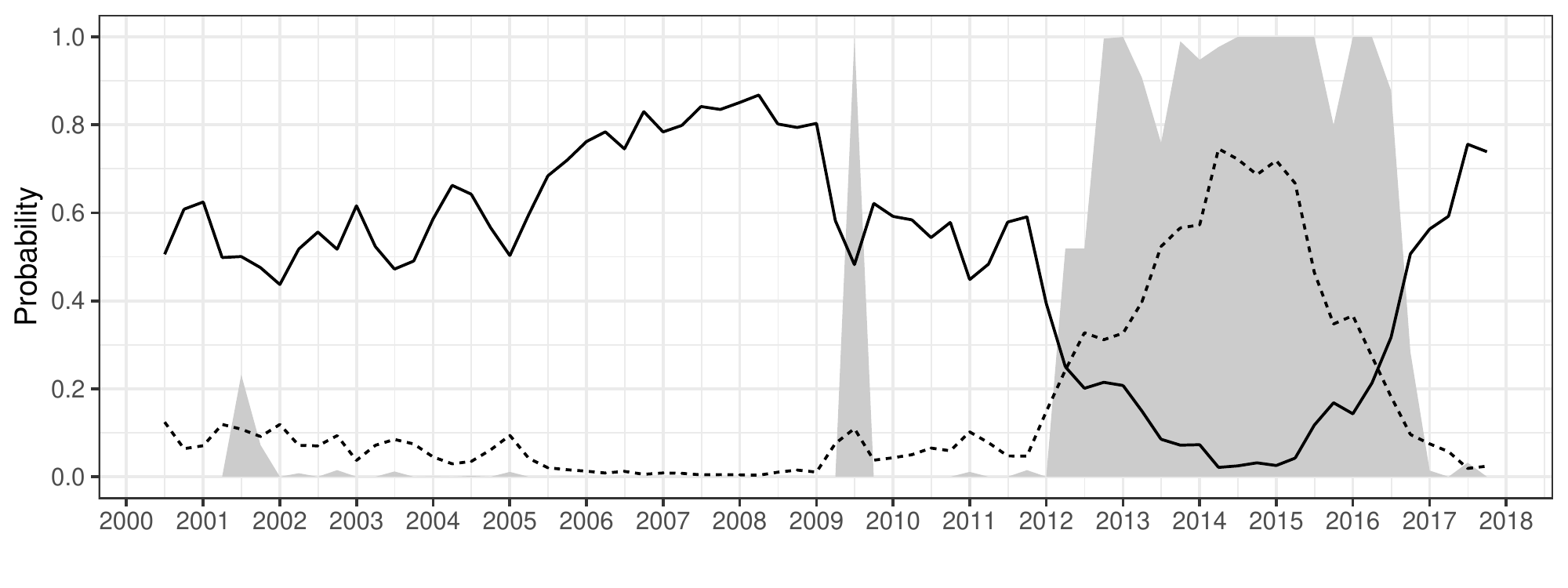}
	\caption{Posterior mean of transition probabilities and filtered probabilities of being in the expansion or recession state.}\label{fig:transition-prob}\vspace{-0.3cm}
	\caption*{\footnotesize\textit{Notes}: The solid line in the upper part indicates $\Pr(S_t=0|S_{t-1}=1)$, the dashed line in the lower part denotes $\Pr(S_t=1|S_{t-1}=0)$. The gray shaded area indicates the posterior mean probability of the recession state.}
\end{figure}

In this section, we provide evidence for time variation in the posterior mean of the transition probabilities. Time-varying transition probabilities alongside posterior regime probabilities are depicted in \autoref{fig:transition-prob}, with gray shaded areas referring to the smoothed probability of being in the recession state. The solid and dashed black lines are the posterior median of the off-diagonal elements in the time-varying transition probability matrix, indicate the probability of entering a recession at time $t$ when being the expansionary state in $t-1$ (that is, $\Pr(S_t=1|S_{t-1}=0)$), and vice versa. 

Our model tracks several periods of economic stress, with three recessionary episodes: First, the burst of the  Dotcom bubble and the 9/11 terrorist attacks are visible. Second, the period surrounding the global financial crisis in 2008/2009 is surrounded by increases in the likelihood of moving into a recession. Finally, the European sovereign debt crisis is captured by our model with a prolonged period of elevated probabilities for $S_t=1$ between 2012 and 2016. After the first crisis in our sample, a period of relative stability emerges, evidenced by a decline in the probability of moving into a recession when being in an expansion until the beginning of 2008. This period ends in 2008 with slightly higher probabilities of entering a recessionary phase, reflecting the beginning of the global financial crisis. After elevated probabilities of recession between 2012 and 2016, the period until the end of our sample is characterized by a steady decline of the probability of switching to the recessionary regime.

\subsection{Do parameters differ across regimes?} \label{sec: relevance}
\begin{figure}[!ht]
\centering
\includegraphics[width=\textwidth]{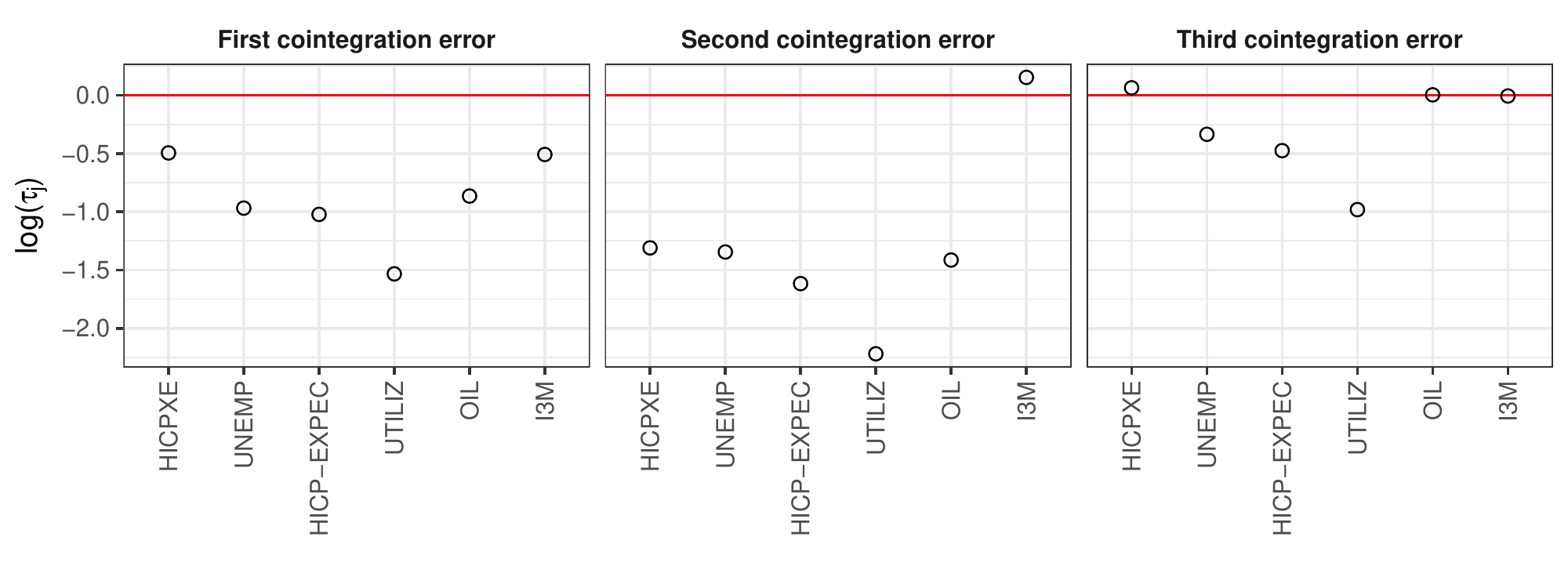}
\caption{Posterior mean of the log of the scaling parameters across equations: Adjustment coefficients associated with the cointegration cointegration errors.}\vspace*{-0.25cm}
\caption*{\footnotesize\textit{Notes}: Variables included are the harmonized index of consumer prices (HICPEX), the unemployment rate (UNEMP), inflation expectations (HICP-EXPEC), Euro area capacity utilization in manufacturing (UTILIZ), oil prices (OIL), and the three-month Euribor (I3M).}
\label{fig: adjustment_scale}
\end{figure}

The key novelty of our proposed approach is that it allows for flexible testing which coefficients differ across regimes. In order to assess differences in parameters, we rely on two visualizations that enable us to assess how much shrinkage is introduced, and to quantify the size of the deviations of $\bm a_{S_t}$ from the common mean $\bm a$. First, we consider the posterior median of the log of $\tau_j$. Figure \ref{fig: adjustment_scale} presents the scaling parameters associated with the short-run adjustment coefficients in $\bm \lambda_{S_t}$ across equations.

\begin{figure}[!ht]
\includegraphics[width=\textwidth]{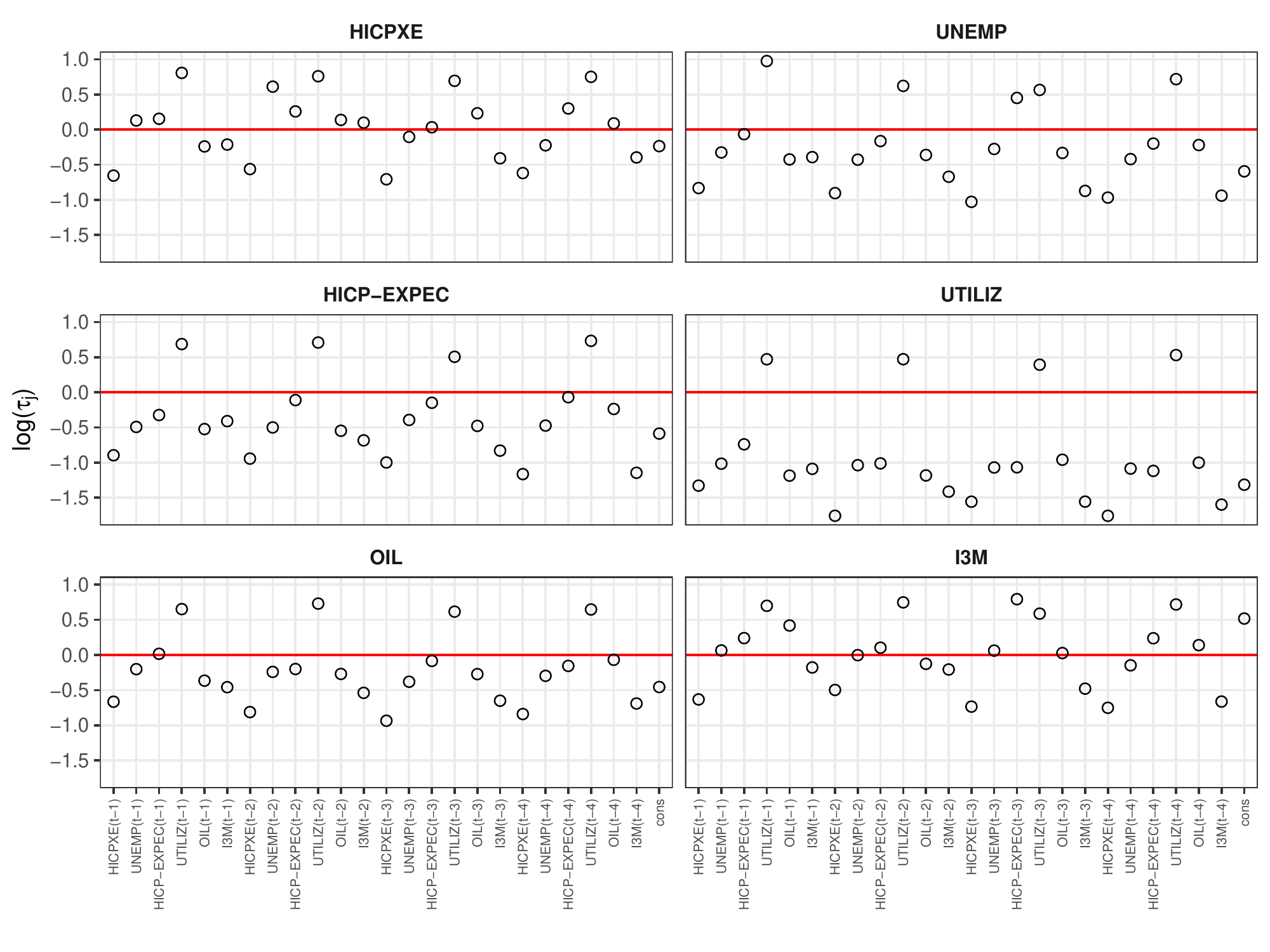}
\caption{Posterior mean of the log of the scaling parameters across equations: autoregressive coefficients.}\vspace*{-0.25cm}
\caption*{\footnotesize\textit{Notes}: Variables included are the harmonized index of consumer prices (HICPEX), the unemployment rate (UNEMP), inflation expectations (HICP-EXPEC), Euro area capacity utilization in manufacturing (UTILIZ), oil prices (OIL), and the three-month Euribor (I3M).}
\label{fig: adjustment_scale_VAR}
\end{figure}

Heterogeneity both across cointegration errors and equations is observable. This suggests that some elements in $\bm\lambda_{S_t}$ are clearly regime dependent, while others are pushed towards homogeneity. For example, the respective coefficients of \textit{UTILIZ} are pushed towards the common mean and thus a linear specification (large negative values on the log-scale), while the coefficients for \textit{HICPXE} and \textit{I3M} differ across regimes. A key takeaway of \autoref{fig: adjustment_scale} is that economic adjustment towards long-run equilibrium values appear to be dependent on the prevailing state of the economy.

We proceed by investigating whether the coefficients associated with the lagged endogenous variables differ for expansions and recessions. \autoref{fig: adjustment_scale_VAR} shows the log scaling parameters for all autoregressive coefficients and the intercept term per equation. A few points are worth noting here. First, an apparent observable pattern is that shrinkage for the autoregressive coefficients tends to be similar across the lags of specific quantities for all equations. In other words, if heavy shrinkage for the first lag of a variable  is introduced, remaining lags are also strongly pushed towards the common mean.

Second, the amount of shrinkage towards linearity is equation specific. The coefficients in the equation for \textit{UTILIZ} in \autoref{fig: adjustment_scale_VAR}, for instance, display only little deviations from the common mean and can thus be treated as time-invariant. This is indicated by large negative values of $\log \tau_j$ for most coefficients.  By contrast, \textit{I3M}, and to a slightly lesser degree \textit{HICPEXE}, appear to require larger breaks in the model coefficients. 

Third, independent of the respective equation, coefficients associated with the lags of \textit{UTILIZ} seem to feature the largest differences between regimes. The lags of \textit{I3M} and \textit{HICPEXE}, on the other hand, are heavily pushed towards the common mean across all equations. Interestingly, this translates to similar patterns of shrinkage on autoregressive coefficients for all variables, even though the amount of shrinkage towards homogeneity differs. Finally, variation in the intercept across regimes is observable for all equations.

\begin{figure}[!ht]
\centering
\includegraphics[width=1\textwidth]{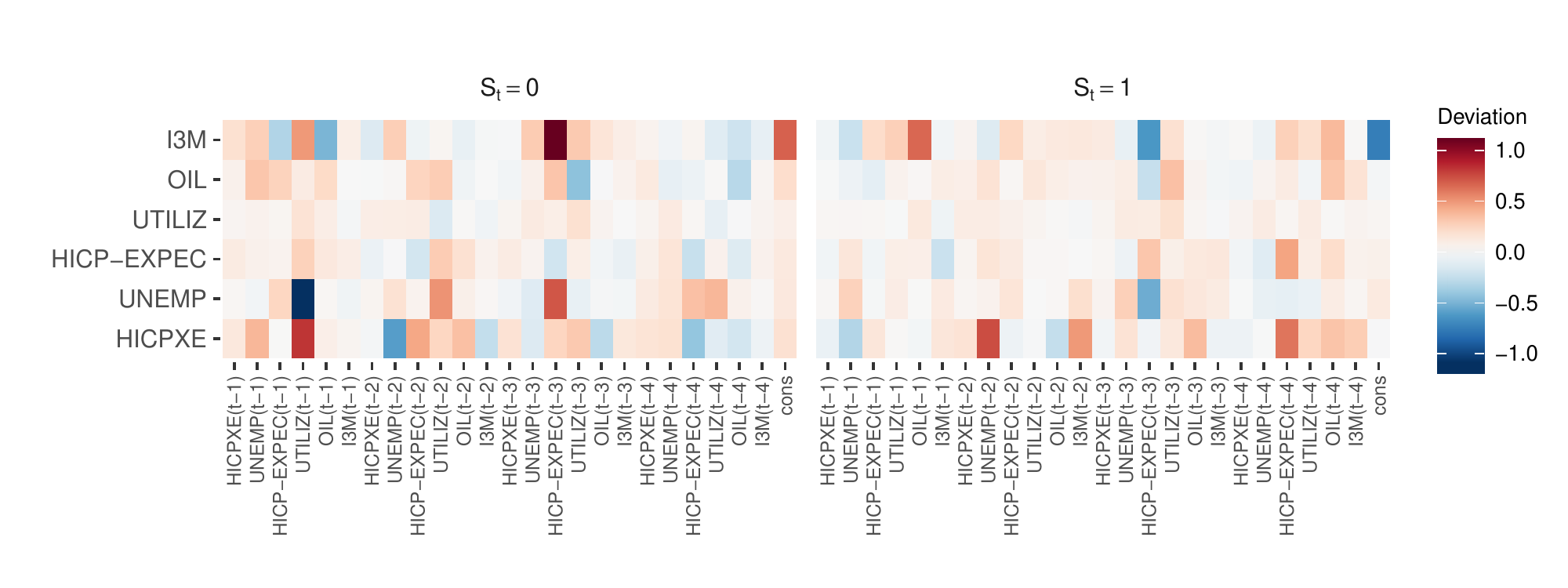}
\caption{Difference of posterior means between state-specific coefficients and the common distribution.}\label{fig:heatmap}\vspace*{-0.25cm}
\caption*{\footnotesize\textit{Notes}: Variables included are the harmonized index of consumer prices (HICPEX), the unemployment rate (UNEMP), inflation expectations (HICP-EXPEC), Euro area capacity utilization in manufacturing (UTILIZ), oil prices (OIL), and the three-month Euribor (I3M).}
\end{figure}

One shortcoming of the analysis presented in Figs. \ref{fig: adjustment_scale} and \ref{fig: adjustment_scale_VAR} is that it is not invariant with respect to the scaling in $\bm{y}_t$ (and its changes). To provide some evidence on the quantitative differences in the autoregressive coefficients, we compute the posterior mean of the distance between regime-specific coefficients and the underlying common distribution. The results are reported in \autoref{fig:heatmap}. We find comparatively large differences for the equations associated with \textit{HICPEXE} and \textit{I3M} in both regimes, in line with the previous discussion of \autoref{fig: adjustment_scale_VAR}. Minor deviations are also apparent in the case of \textit{OIL} and \textit{HICP-EXPEC}, while \textit{UTILIZ} shows only minor differences across regimes except for its own lag. Note that the two states are approximately symmetric, that is, positive deviations in the expansionary regime are typically accompanied by negative differences in the recessionary state. Interestingly, for unemployment \textit{UNEMP} the differences do not mirror each other.

\begin{figure}[!ht]
\includegraphics[width=\textwidth]{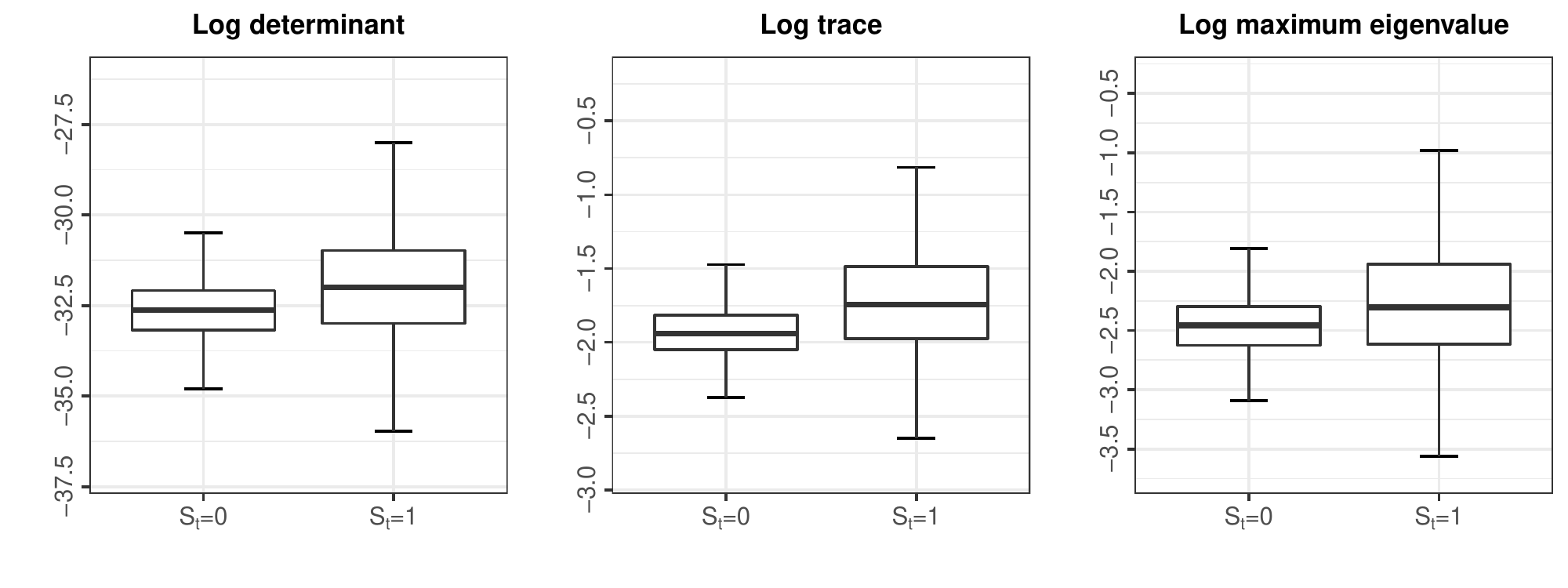}
\caption{Posterior distribution of the log determinant, log trace and the log maximum eigenvalue of the regime-specific variance-covariance matrices}\vspace*{-0.25cm}
\caption*{\footnotesize\textit{Notes}: The thick black line marks the posterior median, the white box covers the 25th and 75th percentile of the posterior distribution. The whiskers refer to the largest (smallest) value within 1.5 times the interquartile range above the 75th percentile (below the 25th percentile).}
\label{fig:logdet}
\end{figure}

Finally, we consider whether the variance-covariance matrices differ across regimes. To this end, \autoref{fig:logdet} presents a boxplot of the marginal posterior distributions of the log determinant, the log trace and the log maximum eigenvalue of the variance-covariance matrix for the recessionary as well as for the expansionary regime. The plots report the median (bold line), together with the first and third quartile and the maximum and minimum value of the posterior distribution of the respective statistics.
Two findings are worth emphasizing. First, considering the posterior median highlights that all considered measures of $\bm{\Sigma}_{1}$ are larger, pointing towards larger error variances in recessions. This is consistent with other findings in the literature that highlight the necessity to allow for heteroscedasticity if recessionary episodes are included in the sample \citep[for forecasting evidence, see,][]{clark2011real}. Second, and finally, we find that posterior uncertainty increases quite sharply during recessionary episodes. This stems from the fact that the recessionary regime features considerably less observations.

To sum up, we find that a subset of the regression coefficients in $\bm a_{0}$ and $\bm a_1$ are different from zero, and we provide strong evidence that error variance-covariance matrices differ across regimes. Our modeling approach stochastically selects a model where only selected VAR coefficients differ while some of short-run adjustment coefficients appear to be regime invariant. Moreover, our findings indicate that the regime allocation is also driven by differences in the variance-covariance matrices and this, to some extent, corroborates findings presented in \cite{sims2006were}. Before proceeding to the forecasting exercise, it is worth emphasizing that our model could be relaxed by introducing separate regime indicators that drive the regime allocation across different regions of the parameter space. However,  our analysis essentially indicates that the bulk of time variation in the parameters of the model stem from the error variance-covariance matrices while the remaining coefficients are strongly pushed towards homogeneity. Introducing separate Markov chains to drive the transition of different blocks of the parameter space would be possible but probably not change much since, in terms of the VAR coefficients, we find comparatively little evidence that parameters tend to change (as opposed to rather strong evidence that the error variances change). This implies that in the recursions used to obtain a draw from the regime indicators, the regime allocation is dominated by the differences in the error variance-covariance matrices. Using different Markov chains to drive the switching behavior of the model would be necessary if we also find that regression coefficients differ sharply across regimes, and thus using a single regime indicator would translate into a miss-specified model. In addition, when it comes to our empirical application we feel that using additional state indicators would render illustrating the key features of the proposed prior much more difficult. We have included an additional discussion in the text where we discuss differences between regimes.

\subsection{Forecasting exercise}\label{sec:forecasting}
We illustrate the merits of our model in a real time out-of-sample forecast exercise. We use the period from 1999:Q1 to 2005:Q4 as an initial estimation period while leaving out the period from 2006Q1 to 2017:Q4 to serve as a verification sample. We then proceed by estimating the model, taking into account different data vintages, and then compute the one-step-ahead predictive distribution. The predictive density is then used to evaluate model performance by computing log predictive scores \citep[LPSs, see][]{geweke2010comparing} relative to a six variable Bayesian VAR model (BVAR) with five lags in \autoref{tab:lps} and a conjugate Minnesota prior in the spirit of \cite{Sims1998}.  Afterwards, we add an additional observation from the holdout sample and re-estimate the model. 

\begin{table*}[h]
\caption{Out-of-sample forecasting results: 2006:Q1 to 2017:Q4.}\vspace*{-1.8em}
\footnotesize
\begin{center}
\begin{threeparttable}
\begin{tabular*}{\textwidth}{@{\extracolsep{\fill}} lrrrrrr}
\toprule
& $r = 1$ & $r = 2$ & $r = 3$ & $r = 4$ & $r = 5$ & r = \textit{NA} \\ 
\midrule 
MS-VECM-TVP  & $14.09$ & $21.69$ & $\bm{39.42}$ & $28.76 $ & $26.59$ & \\
MS-VECM-FTP  & $28.22$ & $29.42$ & $21.71$ & $21.67$ & $29.8$ & \\
VECM   & $22.16$ & $10.02$ & $-66.76$ & $-75.38$& $-117.13$ & \\ 
\midrule
RW & & & & & & $35.89$\\
AR(1) & & & & & & $34.58$\\
\bottomrule
\end{tabular*}
\begin{tablenotes}[para,flushleft]
\footnotesize{\textit{Notes}: The table presents relative log predictive scores for inflation against the BVAR. Large numbers signal better predictive performance. TVP refers to time varying transition probabilities while FTP denote fixed transition probabilities. $r=1$ to $r=5$ refers to the cointegration rank of the VECMs, VECM is a linear VECM, RW and AR(1) denote univariate random walk and autoregressive models of order one.}
\end{tablenotes}
\end{threeparttable}
\end{center}
\label{tab:lps}
\end{table*}

To investigate the merits of our approach, we include additional nested benchmark models.  The set of competing models includes different variants of our model approach, differing in the number of cointegration relations ($r=1,2, \dots, 5$), constant transition probabilities (labeled MS-VECM-FTP), and whether the coefficients feature time-variation (in this case, a linear VECM is adopted). Moreover, to illustrate whether using a multivariate approach pays off in forecasting terms, we include an AR(1) model and the simple random walk (RW) benchmark.

The cumulative LPSs in \autoref{tab:lps} clearly indicate MS-VECM with time varying transition probabilities and $r=3$ as the best performing model in terms of predictive densities. Our highly flexible model does not only outperform the simple BVAR benchmark, but also the simpler random walk specification. In comparison, the linear specification does not perform very well, as it neglects various characteristics the nonlinear model explicitly takes into account. 

\begin{figure}[!t]
\centering
\includegraphics[width=\textwidth]{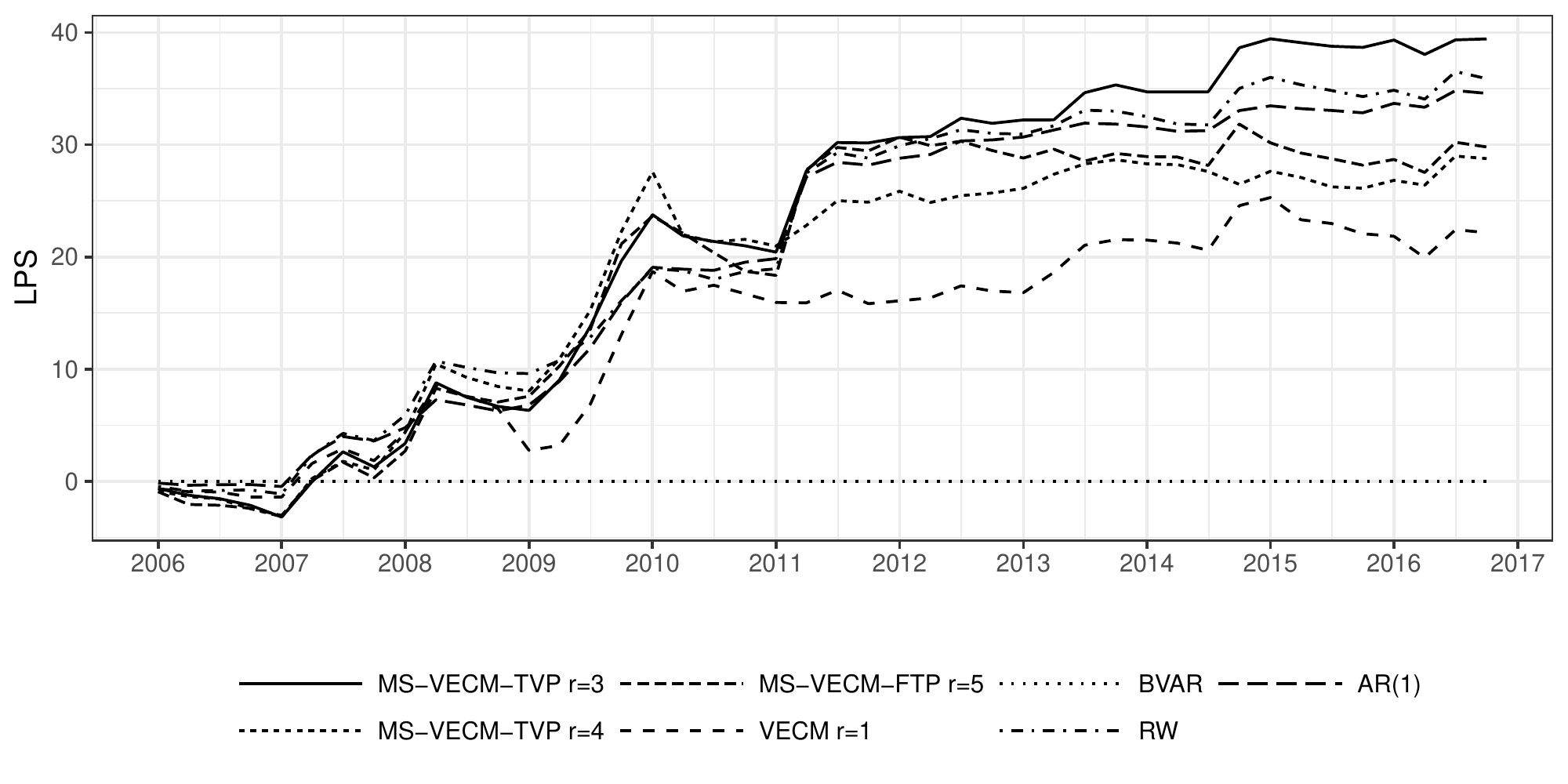}
\caption{Evolution of log predictive scores for inflation relative to the benchmark over the hold-out period: 2006:Q1 to 2017:Q4.}\vspace*{-0.25cm}
\caption*{\footnotesize\textit{Notes}: Large numbers signal better predictive density performance. TVP refers to time varying transition probabilities while FTP denote fixed transition probabilities. $r=1$ to $r=5$ refers to the cointegration rank of the VECMs, RW and AR(1) denote univariate random walk and autoregressive models of order one.}\label{fig:lps_overtime}
\end{figure}

To assess whether differences in forecast performance occur systematically, we assess the evolution of cumulative LPSs relative to the BVAR over time in \autoref{fig:lps_overtime}. While the linear VECM specification underperforms consistently against the nonlinear MS-VECM models throughout the sample, the picture for the remaining models provides a rather homogeneous picture until 2008. Largest relative gains in predictive accuracy for MS-VECM-TVP $r=3$ are observable in 2010 and 2011, and in late 2014.

\section{Closing remarks}\label{sec:concl}
In this paper, we propose a hierarchical nonlinear model that allows for stochastically selecting coefficients that differ across regimes. The proposed MS-VECM discriminates between short- and long-run dynamics and assumes that the transition probability matrix of the underlying Markov process is time-varying.  We assume that the autoregressive coefficients, the error variance-covariance matrices, as well as the short-run adjustment coefficients differ across regimes and arise from a common distribution. Moreover, another novel feature of our model is that the transition distributions are parameterized using a simple binary probit model with the (lagged) cointegration errors included as covariates.

We apply the proposed framework to a medium-scale dataset for the Euro area. Our empirical model discriminates between expansionary and recessionary business cycle stages and allows for assessing whether transition probabilities vary over time. Considering the posterior estimates for the filtered probabilities indicates that our model succeeds in replicating business cycle features of the Euro area. In addition, we investigate which coefficients differ across regimes and how this impacts regime allocation. In-sample evidence is complemented by a real time out-of-sample forecast exercise featuring various benchmark specifications.

Our findings suggest that the proposed approach succeeds in reproducing Euro area recessionary episodes and, additionally, shows that deviations of inflation, unemployment and inflation expectations from their long-run fundamentals drive transitions between regimes. The proposed hierarchical model moreover suggests that only a subset of short-run adjustment coefficients can be assumed to be regime-invariant, and selected autoregressive coefficients differ markedly across expansion and recession periods. A further key modeling aspect is that volatilities of the shocks feature sharp differences across regimes, playing a substantial role in determining regime allocation. The forecast exercise shows that the proposed model is competitive in terms of forecasting Euro area inflation.

\begin{appendices}\crefalias{section}{appsec}
\setcounter{equation}{0}
\renewcommand\theequation{A.\arabic{equation}}
\end{appendices}

\small{\scfont\setstretch{0.85}
\addcontentsline{toc}{section}{References}
\bibliographystyle{bibtex/custom}
\bibliography{bibtex/favar,bibtex/mpShocks,bibtex/pvar,bibtex/gvar,bibtex/ref}}

\end{document}